\documentclass[a4paper,american]{scrartcl}

\usepackage{tikz-cd}
\usepackage{verbatim}
\usepackage{enumerate}
\usepackage{lmodern}
\usepackage[utf8]{inputenc}
\usepackage[T1]{fontenc}
\usepackage{color}
\usepackage{todonotes}
\usepackage{babel}
\usepackage{amsfonts, amsmath, amsthm, amssymb}
\usepackage{setspace}
\usepackage[authoryear]{natbib}
\usepackage[font=small]{quoting}
\usepackage{latexsym}
\usepackage[bookmarks=false,pdfborder={0 0 0},colorlinks=false]{hyperref}
\usepackage{dsfont}
\usepackage{xcolor}
\makeatletter

\usepackage[arrow, matrix, curve]{xy}

\newcommand{\red}[1]{\textcolor{blue}{#1}}

\title{\LARGE How Quantum Mechanics can consistently describe the use of itself}
\author{Dustin Lazarovici\thanks{Université de Lausanne, Faculté des Lettres,		Section de Philosophie. Email: Dustin.Lazarovici@unil.ch} \, and Mario Hubert\thanks{Columbia University, Department of Philosophy. Email: mgh2147@columbia.edu}}

\begin{document}
	\maketitle
\begin{abstract}
	\noindent We discuss the no-go theorem of Frauchiger and Renner based on an ``extended Wigner's friend'' thought experiment which is supposed to show that any single-world interpretation of quantum mechanics leads to inconsistent predictions if it is applicable on all scales. We show that no such inconsistency occurs if one considers a complete description of the physical situation. We  then discuss implications of the thought experiment that have not been clearly addressed in the original paper, including a tension between relativity and nonlocal effects predicted by quantum mechanics. Our discussion applies in particular to Bohmian mechanics. 
\end{abstract}
	
\section{Introduction}	
In a recent Nature paper, Daniela Frauchiger and Renato Renner (2018) present a thought experiment that is meant to show that ''quantum theory cannot consistently describe the use of itself''. \nocite{frauchiger.renner2018} More precisely, the authors formulate three naturally-sounding assumptions and claim that any quantum theory providing a consistent description of the thought experiment must violate at least one of them. 

The first assumption (Q) states the validity of the quantum rule that if the usual measurement formalism predicts an outcome $x=\xi$ with certainty, we can infer that this outcome indeed occurs when the measurement is carried out. It is thereby assumed that the quantum theory can be applied \emph{on all scales}, in particular to Wigner's friend--type experiments, in which a macroscopic system (a laboratory including an experimentalist) is subject to a quantum measurement by another outside observer. A somewhat hidden assumption, included in (Q), is that in a series of such measurements -- e.g. an  experimentalist performs a spin-measurement in her lab, Wigner performs a measurement on the state of this lab -- the first measurement does not lead to a collapse or reduction of the quantum state, i.e. the outside observer can presuppose a linear time evolution of the quantum state until the time of his measurement.   

The second assumption (S) is that the theory be a ``single-world theory'', meaning that measurements have a unique outcome (at least from the viewpoint of the agent carrying out the measurement). The final assumption, (C), is that the quantum theory applied by different agents makes non-contradictory predictions for the experiment.

As natural as these assumptions may sound, the first question we should ask is whether there exists a quantum theory that could make at least a \emph{prima facie} claim to satisfying them all. Going beyond mere instrumentalism, it should be clear that any single-world quantum theory without fundamental collapse law must admit additional variables over and above the wave function or quantum state: If a system (such as a mesasurement device) is in a definite state, despite its quantum state being in a superposition of different (macroscopic) states, it follows that the actual physical state cannot be described by the quantum state alone.



A precise and well worked-out formulation of quantum mechanics that can be applied to the Frauchiger-Renner thought experiment is Bohmian mechanics, also known as the de-Broglie-Bohm theory (see \cite{durr.etal2013} for a modern exposition). Bohmian mechanics is valid on all scales, and the wave function of a closed system always evolves according to a linear Schrödinger equation, so that macroscopic superpositions of quantum states are possible, in principle. However, in Bohmian mechanics, the complete state of a physical system is described by its wave function \emph{and} the actual spatial configuration of particles, which follows a precise law of motion in which the wave function enters. Thus, in the present case, all laboratory equipments, the measuring devices, the brains of the experimentalists, etc., do have a definite configuration at all times, determined by the positions of the particles composing them. 

Our analysis, however, will apply to any single-world quantum theory without fundamental collapse (if there are any other candidates), as long as it assures that \emph{the possible macroscopic configurations registering a measurement outcome correspond to the non-vanishing (in general decoherent) branches of the wave function in the respective pointer basis}. The validity of Born's rule -- which is a theorem rather than an axiom in Bohmian mechanics \cite[ch.\ 2]{durr.etal2013} -- is sufficient for this, though neither the arguments of Frauchiger and Renner nor our response depend on Born statistics holding exactly. 

In the following, we are going to show that any such formulation of quantum mechanics provides a logically consistent description of the thought experiment if the state of the entire system and the effects of all measurements are taken into account. (Our analysis seems more or less in line with the one independently formulated by \cite{sudbery2017}.) Hence, the arguments of Frauchiger and Renner do not reveal any new dilemma (or trilemma) within quantum theory, but rather show the need to apply quantum theory consistently and carefully in these unusual circumstances.


\section{Analysis}
The thought experiment proposed by Frauchiger and Renner consists in a series of four measurements, two of which are standard quantum experiments performed by experimentalist $\bar{F}$ and $F$ in their respective lab $\bar{L}$ and $L$, while the other two are ``Wigner's friend--type'' measurements that outside observers $W$ and $\bar{W}$ perform \emph{on} the labs, which are assumed to have been perfectly isolated up to this point. The experiment starts with a quantum coin-toss: $\bar{F}$ measures a qubit and prepares a spin-1/2-particle depending on the outcome. If the result of the coin-toss is \emph{heads}, she prepares the spin-state $|\downarrow\rangle_S$, if the result is \emph{tails}, she prepares the spin-state $|\rightarrow\rangle_S$. $F$ then receives the so prepared particle and performs a spin-measurement in $z$-direction, obtaining a result $z=\pm \tfrac{1}{2}$. Finally, $\bar{W}$ and $W$ perform a particular quantum measurement on the entire laboratory $\bar{L}$ and $L$, respectively. The precedure is repeated until $\bar{W}$ and $W$  obtain the outcome $\overline{\texttt{ok}}$ and $\texttt{ok}$. The details of these macroscopic measurements are discussed below. \\

           \begin{figure}[ht]
          	\centering \includegraphics[width=0.85\textwidth]{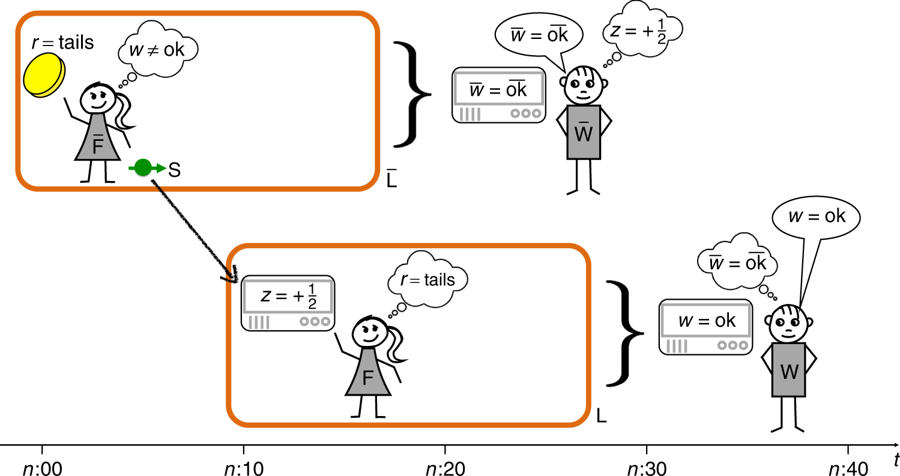}
          	\caption{Sketch of the ``extended Wigner's friend'' thought experiment with the inconsistent conclusions drawn by the experimentalists. Graphic from \citet{frauchiger.renner2018} licensed under CC BY 4.0.}
          \end{figure}
          
\noindent According to the authors, a contradiction arises from the following results, derived by applying  standard quantum rules:

\begin{enumerate}[i)]
	\item$F$ can obtain the result $z=+\frac{1}{2}$ only if the result of the coin toss was \text{tails}.
	\item  $\bar{W}$ can obtain the result $\bar{w}= \overline{\texttt{ok}}$ only if the spin of the particle was \text{up} and $F$ measures $z=+\tfrac{1}{2}$.
	\item  $W$ can obtain the result $w=$ \texttt{ok} only if the result of the coin toss was \text{heads}.
	\item The outcome $\bar{w}=\overline{\texttt{ok}}, w=\texttt{ok}$ has positive probability and will thus occur (almost surely) after sufficiently many trials of the experiment.
\end{enumerate}
 The contradiction is now easily derived. Either the result of the coin toss was \emph{tails}, implying $w=\texttt{fail}$ according to iii). Or the result of the coin toss was \emph{heads}, implying $z=-\frac{1}{2}$ and thus $\bar{w}= \overline{\texttt{fail}}$ according to i) and ii). Hence, the outcome $\bar{w}=\overline{\texttt{ok}}\, \wedge\, w=\texttt{ok}$ is impossible, in contradiction to iv).

 However, a precise analysis of the proposed experiment will reveal that inference iii) is invalid, thus resolving the alleged inconsistency. In brief, the contradiction arises from an insonsistent application of the collapse rule. The derivation of iii)  assumes that $\bar{F}$'s observation of the coin leads to a reduction of her quantum state, while the derivation of ii) assumes that it doesn't. More precisely -- and more in the spirit of Renner's and Frauchiger's argument -- inference iii) is drawn from the point of view of $\bar{F}$ who sees the outcome of the quantum coin toss, prepares either the spin-state $|\downarrow\rangle_S$ or $|\rightarrow\rangle_S$, and thus assumes that she can ignore the part of wave function corresponding to the other possible outcome of the coin toss and the preparation prodcedure she didn't perform. This assumption -- though superficially plausible -- is not a priori warranted by an objective quantum theory such as Bohmian mechanics, and our following analysis will elucidate why it fails in the very particular circumstances of the proposed thought experiment.

While the argument of Frauchiger and Renner is concerned with inferences of agents participating in the experiment (and inferences of agents about the inferences of other agents),  we make a conscious choice not to take these perspectives (at least for now) but describe the experiment in objective terms. Some readers may worry that this misses the point of the Frauchiger-Renner no-go theorem. But then the point of the Frauchiger-Renner no-go theorem is not a good one to begin with. If different ``agents'' make inconsistent predictions by applying a quantum theory that makes a consistent prediction, it can only mean that at least one of the agents applies the theory incorrectly.

  To analyze the scenario correctly, we will consider the wave function of the entire system by tracking the steps of the experiment one by one. For illustrative purposes, we mark in \red{blue} the branches of the wave function that can guide (or in some other sense correspond to) the histories resulting in $\bar{w}=\overline{\texttt{ok}}, w=\texttt{ok}$.
	
\noindent The experiment starts at $t=t_0$ with a ``quantum coin toss'' described by the qubit
\begin{equation}
\sqrt{\frac{1}{3}} |\text{heads}\rangle_{C}  + \red{\sqrt{\frac{2}{3}} |\text{tails}\rangle_{C}}
\end{equation}	
At $t=t_1$, $\bar{F}$ registers the result of the coin toss and prepares the spin-state accordingly:
\begin{equation}\label{eq2}\sqrt{\frac{1}{3}} |\text{heads}\rangle_{C} \otimes|r=\text{heads}\rangle_{\bar{L}}  \otimes|\downarrow\rangle_S + \red{\sqrt{\frac{2}{3}} |\text{tails}\rangle_{C}\otimes|r=\text{tails}\rangle_{\bar{L}}  \otimes|\rightarrow\rangle_S}
\end{equation}
In the following, we will neglect the state of the coin and include it in the state of the lab $\bar{L}$. For ease of notation, we will also drop the symbol for the tensor product.

 At $t_2$, $F$ measures the spin in the z-direction. The resulting state is:
\begin{equation}\label{preparation}\sqrt{\frac{1}{3}}  |\text{heads}\rangle_{\bar{L}}  |z=-\tfrac{1}{2}\rangle_{{L}}+ \red{\sqrt{\frac{1}{3}} |\text{tails}\rangle_{\bar{L}} } \Bigl(|z=-\tfrac{1}{2} \rangle_{{L}} + \red{|z=+\tfrac{1}{2}\rangle_{{L}} } \Bigr),
\end{equation}
where $|z=\pm \tfrac{1}{2} \rangle_{{L}}$ is the state of the laboratory $L$ in which the outcome of the spin-measurement has been recorded. We note that $z(t_2)=+\tfrac{1}{2} \Rightarrow r(t_1)=\textit{tails}$, consistent with the conclusions of Frauchiger and Renner.

 At $t_3$, $\bar{W}$ measures the state of the laboratory $\bar{L}$ with respect to the basis $\Bigl\lbrace|\overline{\mathrm{ok}}\rangle_{\bar{L}}  = \frac{1}{\sqrt{2}}|\text{heads}\rangle_{\bar{L}} - \frac{1}{\sqrt{2}}|\text{tails}\rangle_{\bar{L}}, \, |\overline{\mathrm{fail}}\rangle_{\bar{L}}=\frac{1}{\sqrt{2}}|\text{heads}\rangle_{\bar{L}} + \frac{1}{\sqrt{2}}|\text{tails}\rangle_{\bar{L}}\Bigr\rbrace$. This macroscopic measurement requires a careful treatment. It is described by the following unitary evolution on the level of the quantum state: 
 \begin{subequations}\label{Ameasurement}\begin{align}\nonumber
 |\text{heads}&\rangle_{\bar{L}}|\overline{\text{READY}}\rangle_{\bar{W}}\\
 =&\frac{1}{2}\Bigl(|\text{heads}\rangle_{\bar{L}} + |\text{tails}\rangle_{\bar{L}}\Bigr)|\overline{\text{READY}}\rangle_{\bar{W}} +  \frac{1}{2}\Bigl(|\text{heads}\rangle_{\bar{L}} - |\text{tails}\rangle_{\bar{L}}\Bigr)|\overline{\text{READY}}\rangle_{\bar{W}}\\\nonumber
 \xrightarrow{\text{ S.E. } } &\frac{1}{2}\Bigl(|\text{heads}\rangle_{\bar{L}} + |\text{tails}\rangle_{\bar{L}}\Bigr)|\overline{\text{FAIL}}\rangle_{\bar{W}} +  \frac{1}{2}\Bigl(|\text{heads}\rangle_{\bar{L}} - |\text{tails}\rangle_{\bar{L}}\Bigr)|\overline{\text{OK}}\rangle_{\bar{W}}\\[2ex]\nonumber
 |\text{tails}&\rangle_{\bar{L}}|\overline{\text{READY}}\rangle_{\bar{W}}\\
 =&\frac{1}{2}\Bigl(|\text{heads}\rangle_{\bar{L}} + |\text{tails}\rangle_{\bar{L}}\Bigr)|\overline{\text{READY}}\rangle_{\bar{W}} -  \frac{1}{2}\Bigl(|\text{heads}\rangle_{\bar{L}} - |\text{tails}\rangle_{\bar{L}}\Bigr)|\overline{\text{READY}}\rangle_{\bar{W}}\\\nonumber
 \xrightarrow{\text{ S.E. } } &\frac{1}{2}\Bigl(|\text{heads}\rangle_{\bar{L}} + |\text{tails}\rangle_{\bar{L}}\Bigr)|\overline{\text{FAIL}}\rangle_{\bar{W}} -  \frac{1}{2}\Bigl(|\text{heads}\rangle_{\bar{L}} - |\text{tails}\rangle_{\bar{L}}\Bigr)|\overline{\text{OK}}\rangle_{\bar{W}},
 \end{align}
\end{subequations}

where $|\overline{\text{READY}}\rangle_{\bar{W}}$ denotes the ready-state of $\bar{W}$'s measurement device, S.E. stands for the Schrödinger evolution of the state during the measurement process (coupling of the measured system to the measuring device) and $|\overline{\text{OK}}\rangle_{\bar{W}}, |\overline{\text{FAIL}}\rangle_{\bar{W}}$ denote the ``pointer states'' corresponding to the measurement outcome $\bar{w}= \overline{\texttt{ok}}$ or $\bar{w}= \overline{\texttt{fail}}$, respectively. The result is:
\begin{equation}\begin{split}\label{Ameasurement}
&\frac{1}{\sqrt{12}}\Bigl(|\text{heads}\rangle_{\bar{L}} + |\text{tails}\rangle_{\bar{L}}\Bigr)|\overline{\text{FAIL}}\rangle_{\bar{W}}|-\tfrac{1}{2} \rangle_{\bar{L}} + \frac{1}{\sqrt{12}}\Bigl(|\text{heads}\rangle_{\bar{L}} - |\text{tails}\rangle_{\bar{L}}\Bigr)|\overline{\text{OK}}\rangle_{\bar{W}}|-\tfrac{1}{2} \rangle_{\bar{L}}\\
+&\frac{1}{\sqrt{12}}\Bigl(|\text{heads}\rangle_{\bar{L}} + |\text{tails}\rangle_{\bar{L}}\Bigr)|\overline{\text{FAIL}}\rangle_{\bar{W}}|-\tfrac{1}{2} \rangle_{\bar{L}} - \frac{1}{\sqrt{12}}\Bigl(|\text{heads}\rangle_{\bar{L}} - |\text{tails}\rangle_{\bar{L}}\Bigr)|\overline{\text{OK}}\rangle_{\bar{W}}|-\tfrac{1}{2} \rangle_{\bar{L}}\\
+& \frac{1}{\sqrt{12}}\Bigl(|\text{heads}\rangle_{\bar{L}} + |\text{tails}\rangle_{\bar{L}}\Bigr)|\overline{\text{FAIL}}\rangle_{\bar{W}}|+\tfrac{1}{2} \rangle_{\bar{L}} -  \red{\frac{1}{\sqrt{12}}\Bigl(|\text{heads}\rangle_{\bar{L}} - |\text{tails}\rangle_{\bar{L}}\Bigr)|\overline{\text{OK}}\rangle_{\bar{W}}|+\tfrac{1}{2} \rangle_{\bar{L}}}
\end{split}
\end{equation}
Note that the second and fourth terms cancel (while the first and third terms add up).
We further note that this measurement, performed by $\bar{W}$ on the laboratory $\bar{L}$ of $\bar{F}$, \emph{affects the state of the laboratory}. In particular, if $\bar{L}$ was in an $\lbrace |\text{heads}\rangle, |\text{tails}\rangle \rbrace$--eigenstate prior to $\bar{W}$'s measurement, it will no longer be in such an eigenstate after the measurement was completed. In a Bohm--type theory, this implies, in fact, that the records of $\bar{F}$  (including her brain state and thus her memories) can change due to $\bar{W}$'s interaction. The latter point, however, is not essential to resolving the alleged contradictions. In particular, it remains correct that $\bar{w}(t_3)=\overline{\texttt{ok}} \Rightarrow z(t_2)=+\tfrac{1}{2}$, consistent with the conclusions of Frauchiger and Renner.

Finally, the measurement performed  by $W$ on the laboratory $L$ with respect to the basis $\Bigl\lbrace|\text{ok}\rangle_{L} = \frac{1}{\sqrt{2}}|-\tfrac{1}{2} \rangle_{L} - \frac{1}{\sqrt{2}}|+\tfrac{1}{2} \rangle_{\bar{F}}, \, |\text{fail}\rangle_{L}=\frac{1}{\sqrt{2}} |-\tfrac{1}{2} \rangle_{L} + \frac{1}{\sqrt{2}} |+\tfrac{1}{2} \rangle_{L}\Bigr \rbrace$, is described by the following unitary evolutions:
 \begin{subequations}\label{Wmeasurement}\begin{align}\nonumber
|-\tfrac{1}{2}& \rangle_{L} |\text{READY}\rangle_W \\
= &\frac{1}{2} \Bigl( |-\tfrac{1}{2} \rangle_{L} + |+\tfrac{1}{2} \rangle_{L}\Bigr)|\text{READY}\rangle_W + \frac{1}{2} \Bigl( |-\tfrac{1}{2} \rangle_{L} - |+\tfrac{1}{2} \rangle_{L}\Bigr)|\text{READY}\rangle_W\\\nonumber
\xrightarrow{\text{ S.E. } } &\frac{1}{2} \Bigl( |-\tfrac{1}{2} \rangle_{L} + |+\tfrac{1}{2} \rangle_{L}\Bigr)|\text{FAIL}\rangle_W + \frac{1}{2} \Bigl( |-\tfrac{1}{2} \rangle_{L} - |+\tfrac{1}{2} \rangle_{L}\Bigr)|\text{OK}\rangle_W\\[2ex]\nonumber
|+\tfrac{1}{2}& \rangle_{L} |\text{READY}\rangle_W \\
= &\frac{1}{2} \Bigl( |-\tfrac{1}{2} \rangle_{L} + |+\tfrac{1}{2} \rangle_{L}\Bigr)|\text{READY}\rangle_W - \frac{1}{2} \Bigl( |-\tfrac{1}{2} \rangle_{L} - |-\tfrac{1}{2}\rangle_{L}\Bigr)|\text{READY}\rangle_W\\\nonumber
\xrightarrow{\text{ S.E. } }&\frac{1}{2} \Bigl( |-\tfrac{1}{2} \rangle_{L} + |+\tfrac{1}{2} \rangle_{L}\Bigr)|\text{FAIL}\rangle_W - \tfrac{1}{2} \Bigl( |-\tfrac{1}{2} \rangle_{L} - |-\tfrac{1}{2} \rangle_{L}\Bigr)|\text{OK}\rangle_W
\end{align}\end{subequations}


\noindent The final state for the complete system at $t_4$ (the end of the experiment) is thus:
\begin{equation}\begin{split}\label{final}
&\sqrt{\frac{3}{16}}\Bigl(|\text{heads}\rangle_{\bar{L}} +|\text{tails}\rangle_{\bar{L}}\Bigr) \Bigl( |-\tfrac{1}{2} \rangle_{L} + |+\tfrac{1}{2} \rangle_{L}\Bigr)|\overline{\text{FAIL}}\rangle_{\bar{W}}|\text{FAIL}\rangle_W\\
+&\frac{1}{\sqrt{48}}\Bigl(|\text{heads}\rangle_{\bar{L}} +|\text{tails}\rangle_{\bar{L}}\Bigr) \Bigl( |-\tfrac{1}{2} \rangle_{L} - |+\tfrac{1}{2} \rangle_{L}\Bigr)|\overline{\text{FAIL}}\rangle_{\bar{W}}|\text{OK}\rangle_W\\
-&\frac{1}{\sqrt{48}}\Bigl(|\text{heads}\rangle_{\bar{L}} -|\text{tails}\rangle_{\bar{L}}\Bigr) \Bigl( |-\tfrac{1}{2} \rangle_{L} + |+\tfrac{1}{2} \rangle_{L}\Bigr)|\overline{\text{OK}}\rangle_{\bar{W}}|\text{FAIL}\rangle_W\\
+&\red{\frac{1}{\sqrt{48}}\Bigl(|\text{heads}\rangle_{\bar{L}} -|\text{tails}\rangle_{\bar{L}}\Bigr) \Bigl( |-\tfrac{1}{2} \rangle_{L} - |+\tfrac{1}{2} \rangle_{L}\Bigr)|\overline{\text{OK}}\rangle_{\bar{W}}|\text{OK}\rangle_W} ,
\end{split}\end{equation}
where each summand (after solving all brackets) represents a decoherent branch of the total wave function corresponding to the possible record states. In particular, the ``stopping condition'' $\bar{w}=\overline{\texttt{ok}}, w=\texttt{ok}$ is indeed a possible outcome. Applying the Born rule to the final state \eqref{final}, we see that the probability of this outcome is $4\cdot\frac{1}{48}=\frac{1}{12}$, consistent with the computation of Frauchiger and Renner. 

The crucial observation is now that $w(t_4)= \texttt{ok} \nRightarrow r(t_1) = \textit{heads}$, i.e. inference iii) is unjustified. In fact, the branch containing $|\overline{\text{OK}}\rangle_{\bar{W}}\lvert\text{OK}\rangle_W$ evolved from the last summand in  \eqref{Ameasurement}, which in turn comes from the $|\text{tails}\rangle_{\bar{L}}\otimes|\rightarrow\rangle$ branch in \eqref{eq2}. In Bohmian mechanics, we could actually follow the corresponding quantum flux on configuration space to see that configurations constituting the final outcome $\bar{w}=\overline{\texttt{ok}}, w=\texttt{ok}$ evolve from configurations at $t=t_1$ in which the coin (and the records of $\bar{F}$) showed \emph{tails}.

On the side, we observe again that the measurement of $W$ can change the state of $L$, e.g. from a state in which $z=+\tfrac{1}{2}$ was recorded into a state in which $z=-\tfrac{1}{2}$ is recorded. (That is, even an inference such as $z(t_4)=-\tfrac{1}{2} \Rightarrow z(t_2)=-\tfrac{1}{2}$ would be unjustified.) In fact, as we see from \eqref{final}, all combinations of record states for the four experimentalist are possible at the end of the experiment.

\section{The fallacy}

Why exactly does the inference iii) fail? Frauchiger and Renner assume that when $\bar{F}$ sees the outcome \emph{tails} of the quantum coin toss and prepares the spin state $|\longrightarrow\rangle_S$, she can ignore the first term in \eqref{preparation}, corresponding to the $|\downarrow\rangle$--spin preparation, for the whole remainder of the experiment. In this case, we would indeed conclude that 
\begin{equation*}
|r=\text{tails}\rangle_{\bar{L}} |\longrightarrow\rangle_S \longrightarrow |r=\text{tails}\rangle_{\bar{L}} \frac{1}{\sqrt{2}} \Bigl( |-\tfrac{1}{2} \rangle_{L} + |+\tfrac{1}{2} \rangle_{L}\Bigr) \longrightarrow |\text{FAIL}\rangle_{W}\,. 
\end{equation*}
However, we cannot ignore the ``empty wave'', corresponding to the $|\downarrow\rangle$--spin preparation, since the macroscopic measurement performed by $\bar{W}$ brings the $|\text{heads}\rangle_{\bar{L}}\otimes |z=-\tfrac{1}{2}\rangle_{L}$ and $ |\text{tails}\rangle_{\bar{L}}\otimes |z=-\tfrac{1}{2} \rangle_{L}$ branches back into a \emph{coherent superposition} -- note the first and second line in \eqref{Ameasurement} --, so that the $|\text{heads}\rangle_{\bar{L}} \otimes |z=-\tfrac{1}{2}\rangle_{L}$ branch -- corresponding to the first line in \eqref{Ameasurement} -- influences the subsequent measurement performed by $W$ on  $L$. Indeed, it is precisely the interference between the first and second line in equation \eqref{Ameasurement} which makes it that the surviving branch, consistent with $\bar{w}=\overline{\texttt{ok}}$, is
\begin{equation*}
\red{-  \frac{1}{\sqrt{12}}\Bigl(|\text{heads}\rangle_{\bar{L}} - |\text{tails}\rangle_{\bar{L}}\Bigr)|\overline{\text{OK}}\rangle_{\bar{W}}|+\tfrac{1}{2} \rangle_{L}}
\end{equation*}
rather than
\begin{equation*}
\textcolor{red}{-  \frac{1}{\sqrt{6}}\Bigl(|\text{heads}\rangle_{\bar{L}} - |\text{tails}\rangle_{\bar{L}}\Bigr)|\overline{\text{OK}}\rangle_{\bar{W}}|\text{fail} \rangle_{L}},
\end{equation*}
making the outcome  $\bar{w}=\overline{\texttt{ok}} \, \wedge \,  w=\texttt{ok}$ possible.

What is important to understand -- and what Bohmian mechanics brings out very clearly -- is that a ``single-world interpretation'' of quantum mechanics (without spontaneous collapse) doesn't give us \emph{a priori} license to ignore the parts of the wave function that do not correspond to (or guide) the \emph{actual} state of the system. In the present scenario, the stopping condition $\bar{w}=\overline{\texttt{ok}}, w=\texttt{ok}$ occurs in histories in which the result of the initial coin-toss was \emph{tails}, but the possibility of this outcome is due to the influence of the other (previously decoherent) branch of the wave function, corresponding to $r(t_1)=\text{\emph{heads}}$. An Everettian might say: the outcome $\bar{w}=\overline{\texttt{ok}}, w=\texttt{ok}$ is the result of an interaction of parallel worlds; for a Bohmian, it is explained by the influence of ``empty waves'' that the Wigner's friend measurements bring back into superposition with the ``guiding wave''. Note that what makes Bohmian mechanics a single-world theory is not that these other branches disappear, but that the wave function alone is not a complete specification of the physical state of a system.

In particular, in Bohmian mechanics, there are precise conditions for the \emph{effective collapse}, that is, for when we can forget about empty branches of the wave function  and describe a subsystem by a reduced quantum state that follows an autonomous Schrödinger evolution \cite[see][ch.\ 2]{durr.etal2013}. They invole, in particular, the assumption that the decoherence of macroscopically disjoint states is (for all practical purposes) irreversible -- an assumption that is typically justified, but explicitely undercut by the postulated interaction between $\bar{W}$ and the laboratory $\bar{L}$ (or between $W$ and $L$).

Indeed, what makes the ``extended Wigner's friend'' experiments \emph{practically impossible}, is not the fact that they are measurements on macroscopic systems (it would of course be unproblematic if $\bar{W}$ just peeked into the lab to check whether $\bar{F}$ recorded ``heads'' or ``tails'') but that the particular measurements with respect to the peculiar $\lbrace |\overline{\mathrm{ok}}\rangle, |\overline{\mathrm{fail}}\rangle\rbrace$--basis requires a \emph{reversal of decoherence} of macroscopic wave functions, and thus precise control over a macroscopic number of degrees of freedom, amounting to a violation of the second law of thermodynamics. 

Anyways, the Bohmian analysis for subsystems will reveal that the effective quantum state of coin and particle (at $t_2$, the time of the spin measurement,) is either $|\text{heads}\rangle_{C} |\downarrow\rangle_S$ or  $ |\text{tails}\rangle_{C}|\rightarrow\rangle_S$, while the effective quantum state of the laboratories (at $t_3$, when the first Wigner's friend measurement occurs,) is the entangled state \eqref{preparation}, independent of the preparation procedure that $\bar{L}$ actually performed. And the reason why the latter is not a linear evolution of the former (together with the coupling to the laboratory states) is that $\bar{W}$ and $W$ are supposed to act on macroscopically disjoint branches of the wave function, including the branches that have not guided the subsystems up to this point.

The situation would be notably different if the labs were not perfectly isolated but there was an external record or trace of the measurement results obtained by $\bar{F}$ and $F$ (and be it only in the configuration of air moluecules around their laboratories). By ``external trace'' we mean a configuration that is correlated with the outcomes in $L$ and $\bar{L}$ yet not affected by the measurements of $\bar{W}$ and/or $W$, so that it preserves the decoherence of the laboratory states. In this (more realistic) case, the state of the laboratories \emph{would} be effectively collapsed and inference ii) $  \left( \bar{w}(t_3)=\overline{\texttt{ok}} \Rightarrow z(t_2)=+\tfrac{1}{2}\right )$ rather than iii) incorrect, leading to a different but still self-consistent prediction for the experiment. 

In any case, what \emph{all} precise quantum theories -- including many-worlds and spontaneous collapse theories -- have in common, is that they make precise and objective statements about if and how the wave function / quantum state is collapsed in a particular situation. Frauchiger and Renner assume that this fact is relative to different agents: for $\bar{L}$, the quantum coin toss results in an irreversible collapse, for $\bar{W}$ it doesn't. By this alone, it should not be surprising that they arrive at contradictory conclusions. 

In conclusion, a precise quantum theory such as Bohmian mechanics provides a consistent description of the thought experiment without violating any of the three assumptions (Q), (S) and (C). The no-go theorem stated by Frauchiger and Renner is thus incorrect. To be fully transparent, we cite the exact formulation of assumption (Q) that the authors provide in their paper:
\begin{quote} 
	Suppose that agent $A$ has established that
	\item Statement A(i): ``System $S$ is in state $\lvert\psi\rangle_S$ at time $t_0$.
	
	\item Suppose furthermore that agent $A$ knows that
	
	\item Statement A(ii): ``The value $x$ is obtained by a measurement of $S$ w.r.t. the family $\lbrace \pi^{t_0}_x \rbrace_{x \in X}$ of Heisenberg operators relative to time $t_0$, which is completed at time $t$.''
	
	\item If $\langle \psi \mid  \pi^{t_0}_\xi \mid \psi \rangle =1$ for some $\xi \in X$ then agent $A$ can conclude that
	\item Statement A(iii): ``I am certain that $x=\xi$ at time $t$.''
\end{quote}
We did not formulate our analysis in the Heisenberg picture (because we believe that it only obfuscates the real issues), but it would be straight-forward to do so. In any case, the contradiction derived by Frauchiger and Renner is simply the result of inconsistent and -- in at least one case -- wrong assumptions about what the state of the subsystem in premise A(i) actually is. With the quantum states identified above, the inferences drawn in accord with assumption (Q) are perfectly valid and consistent.

\section{What the thought experiment actually shows}

In sum, we fail to see much of foundational relevance in the no-go theorem formulated by Frauchiger and Renner. The only no-go is to ignore relevant parts of the physical situation, or assume that one and the same subsystem has a different quantum state for different observers. 

Indeed, the quantum theories that are actually embarrassed by the thought experiment are those claiming that the wave function or quantum state represents, in some sense, the epistemic state of an observer, that it is defined in terms of someone's knowledge or information or believe. As our analysis clearly brings out, the sole fact that $\bar{F}$ \emph{knows} the outcome of the quantum coin toss doesn't prevent the other branch of the wave function from being causally efficacious. And the fact that it can be causally efficacious is due to $\bar{W}'s$  measurement affecting the state of the system -- whether $\bar{F}$ \emph{knows} that this measurement will occur, or not. To understand the physics of the thought experiment, we must therefore take the wave function seriously as an objective, physical degree of freedom -- a lesson already emphasized by the celebrated PBR-theorem \citep{pusey.etal2012}.  

There are other aspects of the Frauchiger-Renner thought experiment that we find more interesting and instructive than the problem the authors are trying to construct. For one, as mentioned in our analysis, the macroscopic quantum measurements performed by $\bar{W}$ and $W$ are so invasive that they can change the actual state of the respective laboratory, including the records and memories (brain states) of the experimentalists in it.  Notabene, there is no retroactive change of the past; the states of the laboratories may rather change \emph{post factum} and no longer provide accurate records of their respective history. It is tempting to point to this Orwellian scenario as the solution of the ``paradox''. This, however, would be wrong, since the contradiction alleged by Frauchiger and Renner arises from inferences from and to actual measurement outcomes. Their point is not that the experimentalist will end up with inconsistent records or memories after the experiment was concluded. The observation is nonetheless relevant, as it provides a striking illustration of the fact that a ``quantum measurement'' is not a purely passive process but an invasive interaction that affects the state of the measured system (cf. \citet[ch. 23]{bell2004}).

Another noteworthy feature of the Wigner's friend--type measurements is that they have a \emph{nonlocal} effect: by acting on the system $\bar{L}$, $\bar{W}$ influences the outcome of the measurement performed by $W$ on the system $L$ -- no matter how far these two systems are separated. Nonlocality, of course, is an essential feature of any quantum theory, as brought out conclusively by Bell's theorem \cite[Chs. 2, 16, 24]{bell2004}. It may be useful to note why it cannot be exploited for faster than light signaling in the described experiment. If $W$ knew that the outcome of the quantum coin-toss was \emph{heads}, he could infer from obtaining the result \texttt{ok} that $\bar{W}$ has performed his measurement, even if this measurement occurs at spacelike separation. However, to know the outcome of the coin-toss, $W$ would have to obtain some sort of record (via a regular communication channel). And then, to perform his measurement, $\bar{W}$ would have to act on this record as well (otherwise the state of $\bar{L}$ would remain effectively collapsed), which cannot be done at overlight speed, i.e. at spacelike separation from $W$'s measurement.

Frauchiger and Renner allude to the nonlocality only indirectly, by stating that: 
 
 \begin{quote}
 	[T]he time order of the measurements carried out by agents $\bar{W}$ and $W$ is relevant within Bohmian mechanics. If agent $W$ measured before agent $\bar{W}$ then, according to Bohmian mechanics, statement $\bar{W}^{n:22}$ [our inference ii)] would be invalid whereas $\bar{F}^{n:02}$ [our inference iii)] would hold. This is a clear departure from standard quantum mechanics, where the time order in which agents $\bar{W}$ and $W$ carry out their measurements is irrelevant, because they act on separate systems.
 \end{quote}
 
\noindent The first clarification we should make is that the dependence on the order of measurements follows from the Schrödinger evolution alone. It does not rely in any specific way on the Bohmian ``interpretation''. It is thus doubtful whether it is really true that in ``standard quantum mechanics ... the time order in which agents $\bar{W}$ and $W$ carry out their measurements is irrelevant''. The only principle or theorem we are aware of, is that since the measurements of $W$ and $\bar{W}$ happen on separated systems (and are associated with commuting observables) the \emph{joint probabilities of their outcomes} are independent of any order in which the measurements are performed. This result -- which is also a theorem in Bohmian mechanics -- is not violated here.
 The \emph{actual history} of the experiment, however, can indeed depend on the order in which the measurements  $W$ and $\bar{W}$ are performed. And this fact is indeed problematic \emph{in a relativistic context}, if the two measurements occur at spacelike separation. This, however, is a known issue that has already been demonstrated by the use of ``Hardy's paradox'' \citep{hardy1993}, on which the thought experiment of Frauchiger and Renner is modelled, as well \cite[see][]{berndl.etal1996}. The relevant conclusion to draw here is that, in a single-world, no-collapse quantum theory, quantum equilibrium or Born's rule cannot hold along arbitrary foliations of spacetime. In other words, assumptions (S) and (Q) are incompatible with a (strong) relativity principle: 
 
 \begin{itemize}
 \item[(R)] All Lorentz frames are equivalent in evaluating the predictions of the theory.\end{itemize}
 
\noindent They are not incompatible with formal and empirical Lorentz invariance, as demonstrated by the Lorentz invariant generalizations of Bohmian mechanics discussed by \cite{durr.etal2013a}. A related no-go result, based on Bell's theorem, is proven in \cite{gisin2011}, see also \cite[ch. 5]{albert2015} for an argument showing the tension between relativity and nonlocality on the level of the quantum state. 
The reason why this issue is more commonly addressed in the context of Bohmian mechanics, may be that Bohmian mechanics is simply the only consistent, single-world, linear quantum theory out there. In any case, one virtue of the extended Wigner's friend thought experiment  may be that by magnifying quantum phenomena to macroscopic scales, it shows the untenability of the operationalism behind which the tension between relativity and nonlocality is usually hidden. Obviously, something \emph{actually happens} in the laboratories. A serious quantum theory should be able to describe that, but it cannot do so in a fully relativistic way. This, of course, is not the point that Frauchiger and Renner are making in their paper. We believe that it should be.

\section{On the Bohmian perspective}
Discussing the thought experiment from the point of view of Bohmian mechanics, Frauchiger and Renner make the following statement (in reference to \cite[ch. 2]{durr.etal2013}):
\begin{quote} ``According to the common understanding, Bohmian mechanics is a `theory of the universe' rather than a theory about subsystems. This means that agents who apply the theory must in principle always take an outside perspective on the entire universe, describing themselves as part of it.''\end{quote}
	
\noindent This statement is misleading, although the gist of it is correct. The way in which the statement is correct for Bohmian mechanics, is the way in which it is correct for any serious theory that could be considered as a candidate for a fundamental theory of nature. \emph{Any} fundamental physical theory is first and foremost a theory about the universe as a whole, whose application to subsystems must be justified or derived from an analysis of the fundamental (universal) laws. If we want to apply Newtonian mechanics to a subsystem consisting of a sun and its nearest planet, we must first consider a larger system -- and ultimately the entire universe (the only truly closed system) -- to know whether the influence of other masses can be ignored at the desired level of accuracy. 	Bohmian mechanics allows for a precise analysis ``from the universe to subsystems'', telling us if and why the measurement formalism of textbook quantum mechanics is applicable, and, in particular, if and how a subsystem can be described by its own quantum state. Standard quantum mechanics seems unable to provide such an analysis, since it always has to assume a split -- although a shifty one -- between a measured system and an external observer.
	
Continuing their discussion of Bohmian mechanics, Frauchiger and Renner go on to claim that 
\begin{quote}``This outside perspective is identical for all agents, which ensures consistency and hence the validity of Assumption (C). However, because (S) is satisfied, too, it follows from Theorem 1 that (Q) must be violated.''\end{quote} We disagree. As explained above, all three assumptions (C), (S) and (Q), if properly applied, hold true in Bohmian mechanics. In particular, by taking the ``outside perspectice'', none of the experimentalists needs to come to the conclusion that assumption (Q) is violated. They will rather come to a -- correct -- conclusion about what the right quantum states are to which assumption (Q) refers. 

The contradiction derived by Frauchiger and Renner simply arises from the fact that one experimentalist, $\bar{F}$, uses the wrong quantum state to make predictions about the outcome of a later measurement. Her wrong assumptions may be (subjectively) justified if she doesn't know about the very special procedure that $\bar{W}$ carries out on her laboratory. But this -- to put it bluntly -- is her problem, not a problem with Bohmian mechanics or quantum theory in general. 
	


	
\citet[ch. 12]{durr.etal2013} make an analogy between the wave function in Bohmian mechanics and the Hamiltonian in classical (Hamiltonian) mechanics. The analogy is not a perfect one, but it works for the present purpose. The role of the wave function in Bohmian mechanics is first and foremost to guide the motion of particles. On the fundamental level, there is only one wave function -- the wave function of the universe. The conditional wave function of a particular subsystem is then defined in terms of the universal wave function and the configuration of the environment. In many relevant cases, this conditional wave function is an effective one, allowing for an autonomous description of the subsystem. And in most of these cases, the effective wave function is the one that an experimentalist has actually prepared. (For the difference between conditional and effective wave function, see \citet[ch. 2]{durr.etal2013}.) Similarly, the effective Hamiltonian of a subsystem in classical mechanics depends on the fundamental laws (the Hamiltonian of the ``entire universe'') and the configuration of the environment. In many relevant cases, the effective Hamiltonian will be one that an experimentalist has ``prepared''. But if there are non-negligible influences from outside her subsystem -- let's say Wigner tweaking an electromagnetic field to interfere in a particular way with the one she has produced in her lab -- the effective Hamiltonian will cease to be the one that the experimentalist would assume based on her preparation procedure alone. \emph{Obviously so}, we would add. There are things happening outside her lab that will  the state of the system and thus the outcome of the experiment. 
	
All in all, the mystery behind the Frauchiger-Renner ``paradox'' is not much deeper than that. The main point of distinction of the extended Wigner's friend scenario is that creating interference between fields in physical space is easy, while creating interference between macroscopic wave functions in configuration space is practically impossible. One is a common phenomenon, the other so farfetched that orthodox quantum mechanics got away with ignoring or denying the mere possibility for decades. This is what lends \emph{prima facie} plausibility to the untenable assumptions that Frauchiger and Renner smuggle into their proof. This, and the usual ambiguities of the quantum algorithm, including unwarranted intuitions about a special role of ``agents'' or ``observers''. It is one of the great virtues of Bohmian mechanics that it is not just an algorithm, but a precise theory that can be analyzed for any well specified physical situation. And since the theory can consistently describe nature -- providing a well-defined mathematical law for the motion of particles --, consistently describing ``the use of itself'' is just a matter of applying it correctly. \\

\noindent \textbf{Acknowledgements:} We thank Renato Renner for the very patient and productive discussion. We thank Sheldon Goldstein, Anthony Sudbury, Antoine Tilloy and Serj Aristarkhov for helpful comments. Mario Hubert acknowledges funding from the Swiss National Science Foundation Early Postdoc.Mobility Fellowship, grant no. 174745.
 
\bibliography{renner} 
\bibliographystyle{apalike}

\end{document}